# Room temperature nonlinear Hall effect and wireless RF rectification in Weyl semimetal TaIrTe$_4$

Dushyant Kumar[1], Chuang-Han Hsu[1], Raghav Sharma[1], Tay-Rong Chang[2,3,], Peng Yu[4], Junyong Wang[5,6], Goki Eda[5,6,7], Gengchiau Liang[1,5], and Hyunsoo Yang[1,5*]

[1]Department of Electrical and Computer Engineering, National University of Singapore, 117576, Singapore
[2]Department of Physics, National Cheng Kung University, Tainan 701, Taiwan
[3]Center for Quantum Frontiers of Research & Technology (QFort), National Cheng Kung University, Tainan, 701 Taiwan
[4]School of Materials Science and Engineering, State Key Laboratory of Optoelectronic Materials and Technologies, Sun Yat-sen University, Guangzhou 510275, People's Republic of China
[5]Centre for Advanced 2D Materials, National University of Singapore, 117546, Singapore
[6]Department of Physics, National University of Singapore, 117542 Singapore
[7]Department of Chemistry, National University of Singapore, 117551, Singapore
*e-mail: eleyang@nus.edu.sg

The nonlinear Hall effect (NLHE), which can produce a transverse voltage without any magnetic field, is a potential alternative for rectification or frequency doubling. However, the low temperature detection of NLHE limits its applications. Here, we report the room-temperature NLHE in a type-II Weyl semimetal TaIrTe$_4$, which hosts a robust NLHE due to substantial broken inversion symmetry and large band overlapping at the Fermi level. We also observe a temperature-induced sign inversion of NLHE in TaIrTe$_4$. Our theoretical calculations suggest that the observed sign inversion is a result of temperature-induced shift in the chemical potential indicating a direct correlation of NLHE with the electronic structure at the Fermi surface. Finally, the room-temperature NLHE in TaIrTe$_4$ is exploited to demonstrate the wireless RF rectification with zero external bias and magnetic field. This work opens a door to realizing room temperature applications based on the NLHE in Weyl semimetals.

The Hall effects, because of their deep connections with the geometrical phases and topology, have been one of the main research focuses in modern materials and physics, leading to the search of topological states of matter and many practical applications[1–3]. The Hall conductivity, in the linear response regime, requires time-reversal symmetry to be broken, either by the magnetization or external magnetic field[3,4]. However, a recent study predicted that in the non-linear response regime, the Hall conductivity can still survive in the presence of time-reversal symmetry but with broken inversion symmetry[5]. This is called the nonlinear Hall effect (NLHE). The NLHE is governed by the $k$-space Berry curvature dipole (BCD)[5], an intrinsic property related to the geometrical characteristic of electronic states. In particular, the second-order nonlinearity exhibits two components; one is the voltage oscillating at twice the frequency $2\omega$ of the driving alternating current and the other is a d.c. component that is generated due to the rectification effect[5–9] (Supplementary Notes 1 and 2). Therefore, the NLHE can be utilized in applications requiring frequency doubling or rectification such as energy harvesting, wireless communications and IR detectors[6,10]. However, it requires a room temperature detection of NLHE. Despite of several studies of NLHE in topological materials and 2D materials, the room-temperature NLHE has not yet been achieved[2,11–16]. Considering the Berry phase effect and its possible applications[1], it is of great interest to study nonlinear electrical and optical properties of the newly emerging topological materials especially at room temperature[1,17–19].

Since the Berry curvature is essential to the NLHE, which also constitutes the topology of bands, it motivates to explore topological materials for a large NLHE. Among various topological materials, the time-reversal invariant Weyl semimetals (T-WSM)[20,21] are expected to host large NLHE in two perspectives. First, substantial Berry curvature is likely to be found around Weyl points[12,16], where bands cross at $k$-points in the momentum space and behave as sources or sinks of the Berry curvature[1]. Second, the inversion-symmetry breaking is demanded by both T-WSM and NLHE. In a recent report of NLHE in few-layer WTe2, the origin of sizable BCD was shown to be the interplay between the spin-orbit coupling (SOC) and inversion-symmetry breaking[2,14]. Furthermore, the large nonlinear photoresponse observed in TaIrTe4 has been related to the divergence of Berry curvature[22]. These studies suggest that a T-WSM with substantial SOC and Berry curvature is a potential candidate for a large NLHE.

In addition, it is important to mention that the NLHE is a Fermi liquid property[5], and thus is closely related to the Fermi surface. It was reported that the chemical potential in WTe2 can evolve by several tens of meV by increasing temperature from 2 to 300 K resulting in dramatic changes of the Fermi surface[23]. Therefore, the NLHE in T-WSM is likely to be sensitive to temperature, which may lead to dramatic changes of NLH-signal. Although the NLHE has recently been reported in a few layers $T_d$-WTe2, the study is limited below 100 K and no significant change of NLH-voltage has been observed with temperature[14,15]. The study under a wide range of temperature is still lacking and the room-temperature NLHE, which would be desirable for applications, is not yet reported.

In this work, we study the NLHE in a type-II T-WSM TaIrTe4, which hosts the minimal number of well separated Weyl points with broken inversion symmetry[24–26]. The NLHE is present only when the electric field is applied along the axis of the mirror plane ($\hat{a}$ axis of the crystal) and a sizeable NLHE persists up to room temperature. The sign of the NLH-voltage is observed to change at ~175 K. Our theoretical calculations of BCD in TaIrTe4 demonstrate that by gradually increasing the chemical potential $\mu_E$, the corresponding BCD changes the sign, which verifies the susceptible nature of NLHE to the Fermi surface. The thickness dependence of NLHE is further investigated to validate the inherent crystal symmetry constrains for NLHE when the bulk property of TaIrTe4 is considered. In the end, we demonstrate that the NLHE in TaIrTe4 can be utilized for battery-free energy harvesting applications. This is for the first time, to our knowledge, a room-temperature NLHE has been reported in any material and further utilized for a potential application.

TaIrTe4, a simplest type II T-WSM[24,25], is known to show two bulk structural phases: monoclinic $1T'$ and orthorhombic $T_d$. While the $1T'$ phase belongs to centrosymmetric space group $P2_1/m$, the $T_d$ phase shows non-centrosymmetric space group $Pmn2_1$ (refs.[27,28]). Recent angle-resolved polarized Raman spectroscopy study confirms the absence of inversion symmetry in $T_d$ phase of TaIrTe4, a prerequisite for the NLHE to exist[27]. Considering the symmetry constraints (summarized in Supplementary Note 4) in $T_d$-TaIrTe4, an a.c. electric field ($E_{\hat{a}}^{\omega}$) along the $\hat{a}$ crystal direction should produce a nonlinear Hall current $J_{\hat{b}}^{NLHE} \propto D_{\hat{a}\hat{c}}(E_{\hat{a}}^{\omega})^2$ in $\hat{b}$, driven by BCD ($D_{\hat{a}\hat{c}}$) without any magnetic field.

We fabricate our devices by mechanically exfoliating a single crystal of $T_d$-TaIrTe4 (synthesized by solid state reaction) onto a highly resistive oxidized Si wafer. Rectangular flakes of TaIrTe4 were chosen and patterned along crystallographic directions. Figure 1a shows the Hall bar device geometry and the measurement configurations. Optical images of the Hall bar devices are shown in Fig. 1b. One can clearly see the light blue colored patterned TaIrTe4 flake in the devices with electrodes aligned in crystallographic directions. To confirm the broken inversion symmetry and the crystallographic alignment (see Fig. 1c) with the patterned electrodes, we performed Raman spectroscopy on these devices. A representative Raman spectrum is shown in Supplementary Fig. 1a. The Raman peaks correspond well with the previous report[27]. It is worth noting that two Raman modes at ~110 cm$^{-1}$ and 134 cm$^{-1}$ are regarded as a signature of broken inversion symmetry in TaIrTe4. The crystallographic alignment is confirmed using angle-resolved polarized Raman spectroscopy (Supplementary Note 5). The resistivity measured in different directions (see inset of Supplementary Fig. 1d) shows a strong in-plane electrical anisotropy with $\hat{a}$ being the most conductive direction in agreement with previous studies[27]. We observe $\hat{a}$ to be ~ 4 times more conducting than $\hat{b}$. Our devices also show the linear dependence of current-voltage (IV) characteristics confirming the formation of ohmic electrical contacts (Supplementary Note 6).

In order to study the NLHE, we apply an a.c. current $I^{\omega}$ ($\omega$ = 13.7 Hz), and measure both the first-harmonic $\omega$ and second-harmonic $2\omega$ frequencies of longitudinal and transverse voltages simultaneously using lock-in amplifiers. The second-harmonic transverse voltage corresponds to the NLH-voltage, which is denoted by $V^{2\omega}_{\hat{\beta}-\hat{\alpha}\hat{\alpha}}$, where $\hat{\alpha}$ is the current direction and the voltage is measured along in-plane transverse direction $\hat{\beta}$ ($\hat{\beta} \perp \hat{\alpha}$). First, we study the behavior of NLHE at low temperature for both of the crystallographic axis, $\hat{a}$ and $\hat{b}$ ($\hat{a} \perp \hat{b}$). The experiments are performed on a 20 nm thick Hall bar device at 2 K with zero magnetic field. With current along $\hat{a}$, the first-harmonic longitudinal voltage $V^{\omega}_{\hat{a}-\hat{a}\hat{a}}$ increases linearly with $I^{\omega}$, whereas the first harmonic transverse voltage $V^{\omega}_{\hat{b}-\hat{a}\hat{a}}$ remains small as expected in a Hall bar geometry (Supplementary Note 6). We observe a clear signal for the second-harmonic transverse voltage $V^{2\omega}_{\hat{b}-\hat{a}\hat{a}}$ that is much larger than the longitudinal response $V^{2\omega}_{\hat{a}-\hat{a}\hat{a}}$ (Fig. 1d). It scales quadratically with $I^{\omega}$ i.e. $V^{2\omega}_{\hat{b}-\hat{a}\hat{a}} \propto (I^{\omega})^2$. These results indicate the presence of second-order nature of NLHE in $T_d$-TaIrTe$_4$.

To further confirm its existence, we perform the similar measurements with the current applied along $\hat{b}$. In this case, there should not be any NLH-voltage because of the mirror symmetry $M_{\hat{a}}$, which reflects coordinates along $\hat{a}$. Indeed, as seen from Fig. 1e, the second-harmonic voltage $V^{2\omega}_{\hat{a}-\hat{b}\hat{b}}$ is almost zero with current along $\hat{b}$. $V^{2\omega}_{\hat{b}-\hat{b}\hat{b}}$ is also zero as expected when the external magnetic field is absent. Next, we perform frequency dependent measurements by changing the frequency from 13.7 Hz to 213.7 Hz, keeping the applied current along $\hat{a}$. The output NLH-voltage is consistent and no frequency dependent change is observed (Supplementary Note 6). The frequency independent response validates the essential property of BCD and excludes measurement artifacts such as spurious capacitive coupling. These results confirm the NLHE observed in $T_d$-TaIrTe$_4$ at 2 K, which is stronger than the one in $T_d$-WTe$_2$ (Supplementary Note 7).

We then study the temperature dependent NLHE in TaIrTe$_4$. The NLH-voltage with currents along $\hat{b}$ ($V^{2\omega}_{\hat{a}-\hat{b}\hat{b}}$) stays zero from 2 K up to room temperature (Fig. 1f). However, as shown in Fig. 1g, the NLH-voltage with the current along $\hat{a}$ ($V^{2\omega}_{\hat{b}-\hat{a}\hat{a}}$) gradually decreases with increasing temperature up to ~ 150 K. Interestingly, with further increase in temperature, the sign of $V^{2\omega}_{\hat{b}-\hat{a}\hat{a}}$ inverts and the magnitude increases up to 300 K. We have performed similar measurements in several devices with different geometries (Supplementary Notes 9 and 10). The results are consistent in all devices, which confirms the first observation of NLHE at room temperature.

As mentioned earlier, the chemical potential $\mu_E$ of a T-WSM is likely to be tuned by changing the temperature $T$, and it is possible that the temperature-driven sign inversion of $V^{2\omega}_{\hat{b}-\hat{a}\hat{a}}$ results from the variation in the Fermi surface. To further confirm this scenario, we perform theoretical calculations of BCD with 30 $T_d$-TaIrTe$_4$ monolayers. The model Hamiltonian of the $T_d$-TaIrTe$_4$ slab is extracted from a first-principles tight-binding Hamiltonian of bulk $T_d$-TaIrTe$_4$ (ref.[26]). To verity the corresponding $\mu_E$ at different $T$, we first compare the experimental and calculated carrier density of hole $n_h$ and electron $n_e$, which are shown in Fig. 2a and 2b, respectively, using the method in ref.[29]. At $T$ = 2 K, the carrier density of hole and electron is almost compensated ($n_h = n_e$), which corresponds to $\mu_E = -0.014$ eV in our first-principles tight-binding Hamiltonian. Although the calculated $n_h/n_e$ has a discrepancy when comparing to the experimental result, an approximate estimation shows that the total shift of $\mu_E$ is ~ 0.054 eV from 2 to 300 K.

The calculated BCD response $D_{\hat{a}\hat{c}}$ is shown in Fig. 2c, where the dashed line indicates $\mu_E$

= −0.014 eV. It can be seen in Fig. 2c that the sign of $D_{\hat{a}\hat{c}}$ changes while $\mu_E$ increasing from −0.014 eV to +0.04 eV. This result qualitatively agrees with the experimental observation shown in Fig. 2d. The calculated k-resolve distributions of $D_{\hat{a}\hat{c}}$ at $\mu_E$ = −0.014 eV (for $D_{\hat{a}\hat{c}}$ > 0) and $\mu_E$ = +0.03 eV (for $D_{\hat{a}\hat{c}}$ < 0) are shown in Fig. 2e and 2f, respectively. The green lines are the constant energy contours and the sign of $D_{\hat{a}\hat{c}}$ is indicated by red (> 0) and blue (< 0) colors. One can easily observe that the Fermi surface contours change considerably at different $\mu_E$ in Fig. 2e and 2f. The sign change can be attributed to the electron pockets residing at two sides of the zone center Γ ($k_a = k_b$ = 0). Further calculations for bilayer $T_d$-WTe$_2$ and $T_d$-TaIrTe$_4$ suggest that the carrier density ($n_c$) could play a crucial role for $D_{\hat{a}\hat{c}}$. When comparing the calculated $n_c$ of bilayer $T_d$-WTe$_2$ and $T_d$-TaIrTe$_4$ (Supplementary Note 8), one can find that $n_c$ of $T_d$-TaIrTe$_4$ is almost one order larger than that of $T_d$-WTe$_2$ at $\mu_E$ of charge neutrality, which also agrees with the experimental measurement. Meanwhile, the calculated $D_{\hat{a}\hat{c}}$ of $T_d$-TaIrTe$_4$ at the same $\mu_E$ is also almost one order larger than the one of $T_d$-WTe$_2$, which implies a proportional relationship between $n_c$ and $D_{\hat{a}\hat{c}}$.

In order to further investigate the characteristics of NLHE, we perform similar measurements on devices with different thicknesses (Supplementary Note 9), and extract the strength χ of NLHE as $V^{2\omega}_{\hat{b}-\hat{a}\hat{a}} = \chi (J^{\omega}_s)^2$, where $J^{\omega}_s$ is the surface current density. As depicted in Fig. 3a, a slab can be decomposed into surface and bulk parts. At the middle part of a slab, the translation symmetry is gradually restored and thus its physical properties are dictated by all the bulk symmetries that forbid $D_{\hat{a}\hat{c}}$. On the other hand, the absence of $\overline{M}_{\hat{b}}$ on the surface allows $D_{\hat{a}\hat{c}}$ contributing to the NLHE. Since the NLH-signal can only come from the surface, the decrease of NLH-strength with increasing the thickness is expected in TaIrTe$_4$ slabs. Our experimental results in Fig. 3b indeed demonstrate the expected tendency that the NLH-voltage decreases while the slab thickness increases.

We also examine the scaling law of NLHE in TaIrTe4, which can shed light on the different sources of the observed NLH-signal (Supplementary Note 11). Results suggest the contribution from both the intrinsic Berry curvature and the extrinsic disorder scatterings. Scaling is linear with the conductivity along $a$ ($\sigma_{\hat{a}}^2$) in highly resistive samples, however, as the conductivity increases the scaling law becomes parabolic in consistence with the theoretical prediction[30]. Scaling analysis also suggests that both intrinsic and extrinsic effects could result in the sign inversion of NLH-voltage, however, it cannot be unveiled which one being dominated.

We now exploit the observed room-temperature NLHE in TaIrTe4 to demonstrate the wireless RF rectification with zero external bias (battery free) and magnetic field. The NLHE, being a second order phenomenon, can provide an alternative approach to harvest electromagnetic energy by converting an oscillating electromagnetic field into a d.c. current (Supplementary Notes 1 and 2). The schematic of the rectification experiment is shown in Fig. 4a where the Hall bar device of TaIrTe4 is exposed to electromagnetic waves using a patch microstrip antenna with the central frequency ~2.4 GHz. The cutoff frequency (defined as -3 dB point) of our device is ~5 GHz, high enough to cover the Wi-Fi channel of 2.4 GHz. Following the same convention as for second harmonic measurements of NLHE, we first align the incident wireless electric field $E_{\hat{a}}^{\omega}$ along $\hat{a}$, and measure the rectified d.c. voltage $V_{\hat{b}}^{DC}$ in the transverse direction $\hat{b}$. A clear rectified signal can be seen in Fig. 4b. We also measure the d.c. voltage along $\hat{a}$ ($V_{\hat{a}}^{DC}$) while applying $E^{\omega}$ along the $\hat{b}$ direction, however, as expected from the NLHE symmetry constraints, the $V_{\hat{a}}^{DC}$ remains negligible (Fig. 4c). Rectification measurements with the electric field along $\hat{c}$ (Supplementary Fig. 9) also follow the NLHE symmetry constraints. Furthermore, the rectified d.c. signal in a thicker (96 nm) sample of TaIrTe4 is very small (Supplementary Fig. 10), in consistent with our thickness dependence in Fig. 3.

It is worthwhile to compare our current observation to the previously revealed broadband photocurrent on TaIrTe$_4$ (ref. [22,31]) in the several hundred terahertz regime. First, the phenomenon reported here does not accompany with obvious photoexcitation, and the incident electromagnetic (EM) wave can be regarded as an a.c. electric field[10]. Second, recently studied photocurrents[22,31] mainly result from the third-order response that survives in bulk TaIrTe$_4$, whereas our study emphasizes on the second-order response that can be attributed to the surface contribution. To further validate these results, we perform similar measurements on $T_d$-WTe$_2$, $1T'$-MoTe$_2$ and Pt devices (Supplementary Note 12). As expected, WTe$_2$ with broken inversion symmetry shows some rectified voltages, however the signal is very small, in line with the second harmonic measurements on $T_d$-WTe$_2$ at room temperature (Supplementary Note 7). $1T'$-MoTe$_2$ and Pt, both being centrosymmetric materials, do not show any rectification revealing the essential role of NLHE behind the observed rectification in TaIrTe$_4$. These results show the proof-of-concept of utilizing the NLHE in the energy harvesting and microwave detection applications with specially designed on-chip circuitry for enhancing the rectified voltage.

type-II Weyl semimetal candidate TaIrTe$_4$. *ACS nano* **12,** 4055-4061 (2018).

**Methods**

**Sample Growth.** All the used elements in the growth of TaIrTe$_4$ were stored and acquired in argon-filled glovebox with moisture and oxygen levels less than 0.1 ppm, and all manipulations were carried out in a glovebox. TaIrTe$_4$ single crystals were synthesized by solid state reaction with the help of Te flux. The elements of Ta powder (99.99 %), Ir powder (99.999 %), and Te lump (99.999 %) with an atomic ratio of Ta/Ir/Te = 1 : 1 : 12, purchased from Sigma-Aldrich, were loaded in a quartz tube and then flame-sealed under high-vacuum of $10^{-6}$ torr. The quartz tube was placed in a tube furnace, slowly heated up to 1000 °C and held for 100 h, and then allowed to cool to 600 °C at a rate of 0.8 °C/h, followed by cooling down to room temperature. Shiny, needle-shaped TaIrTe$_4$ single crystals can be obtained from the product. $T_d$-WTe$_2$ and $1T'$-MoTe$_2$ single crystals were purchased from HQ Graphene.

**Raman spectroscopy.** Raman Spectra were measured in a NT-MDT Spectroscopy system. The excitation laser is 532 nm for polarized Raman spectra and 633 nm for non-polarized Raman spectra. The linear polarized laser is focused on samples by a 100× objective lens with a NA of 0.9. The scattered Raman signals go through the same lens and back to a spectrometer. Polarized Raman spectroscopy in the parallel configuration is typically used to precisely determine the crystal orientation of the anisotropic 2D flakes. In this configuration, the analyzer was in front of the entrance slit of the spectrometer in the parallel direction relative to the incident laser. The samples were rotated by a stage with a step of 30 degree keeping all other parameters identical during the measurements.

**Device Fabrication.** TaIrTe$_4$ samples were mechanically exfoliated from a single crystalline bulk $T_d$-TaIrTe$_4$ on to Si/SiO$_2$ (300 nm) substrates. To protect from oxidation and further degradation, a 5 nm thick capping layer of SiO$_2$ was deposited using sputtering. The thickness of TaIrTe$_4$ flakes was first estimated by optical contrast, and then measured by atomic force microscopy (AFM). Standard Hall bar geometry as well as multi-terminal star-like device geometry were patterned by conventional electron beam lithography process followed by Ar-ion etching. Contact electrodes of Ta (5 nm)/Cu (70 nm) were deposited using sputtering.

**Electrical measurements.** The electrical measurements were carried out in the physical property measurement system (PPMS, Quantum Design). An a.c. electric current from a Keithley 6221 source with a frequency of 13.7 Hz was applied to the devices. Fundamental and second-harmonic frequencies of both longitudinal and transverse voltages were measured simultaneously using four Stanford Research SR830 lock-in amplifiers.

**RF rectification measurements.** For the rectification measurement with the experimental geometry shown in Fig. 4a, we use a TM10-mode patch microstrip antenna, which is low cost and easy-fabricated antenna designed on a circuit board. The patch antenna dimensions are chosen for the 2.4 GHz operation with -20 dBm return loss, whose bandwidth is characterized by high return loss of -10 dBm ranging from 2.35 GHz to 2.45 GHz. The patch antenna is designed for vertical polarization of the electric field with TM10-mode propagated in the space. The antenna is fed through an Agilent E8257D signal generator. The ambient Wi-Fi power density irradiated onto the sample is calculated as ~ 8 mW/cm$^2$ for the d.c. response shown in the Fig. 4b. The d.c. voltage was measured in a Keithley 2182A nanovoltmeter. The sample was placed parallel to the antenna such that the incident electric field is along the $\hat{a}$ direction of the Hall bar device, which is taken as 0° for the convention. The sample is rotated using high-precision angular mount.

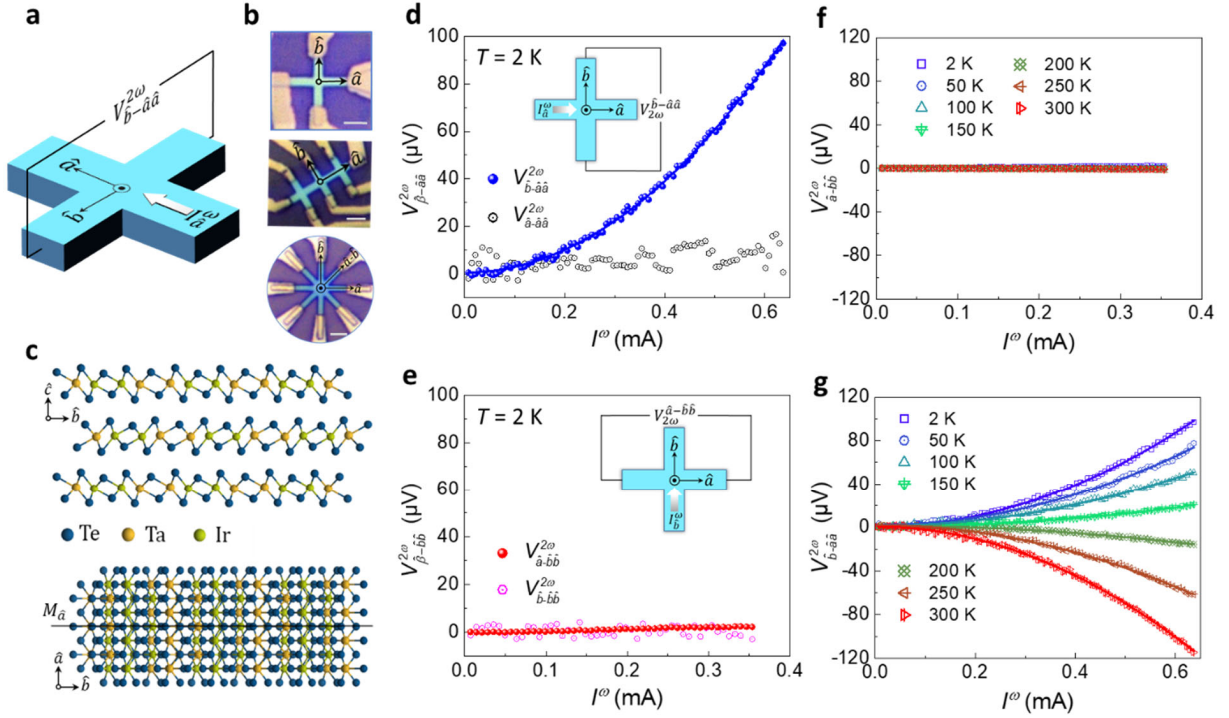

**Fig. 1 | Nonlinear Hall effect in a 20 nm thick Hall bar device of TaIrTe$_4$. a,** Schematic of a Hall bar device and measurement configuration. **b,** Optical images of Hall bars devices. The scale bars are 3 μm **c,** Crystal structure of TaIrTe$_4$ for the side (upper) and top (lower) views. **d,e,** Second-harmonic longitudinal $V^{2\omega}_{\hat{a}-\hat{a}\hat{a}}$ ($V^{2\omega}_{\hat{b}-\hat{b}\hat{b}}$) and transverse $V^{2\omega}_{\hat{b}-\hat{a}\hat{a}}$ ($V^{2\omega}_{\hat{a}-\hat{b}\hat{b}}$) voltages in response to an applied a.c. current $I^{\omega}$ along $\hat{a}$ ($\hat{b}$). Nonlinear Hall voltage (NLH) $V^{2\omega}_{\hat{b}-\hat{a}\hat{a}}$ in **d**, varies quadratically with $I^{\omega}$ when it is applied along $\hat{a}$. However, the NLH-voltage $V^{2\omega}_{\hat{a}-\hat{b}\hat{b}}$ in **e**, is almost zero with $I^{\omega}$ along $\hat{b}$. **f,g,** Temperature evolution of $V^{2\omega}_{\hat{a}-\hat{b}\hat{b}}$ and $V^{2\omega}_{\hat{b}-\hat{a}\hat{a}}$. $V^{2\omega}_{\hat{a}-\hat{b}\hat{b}}$ in **f** is negligible throughout the temperature range of 2 to 300 K, however, $V^{2\omega}_{\hat{b}-\hat{a}\hat{a}}$ in **g** shows a sign change at ~175 K. Solid lines in **d** and **g** are quadratic fit to the data.

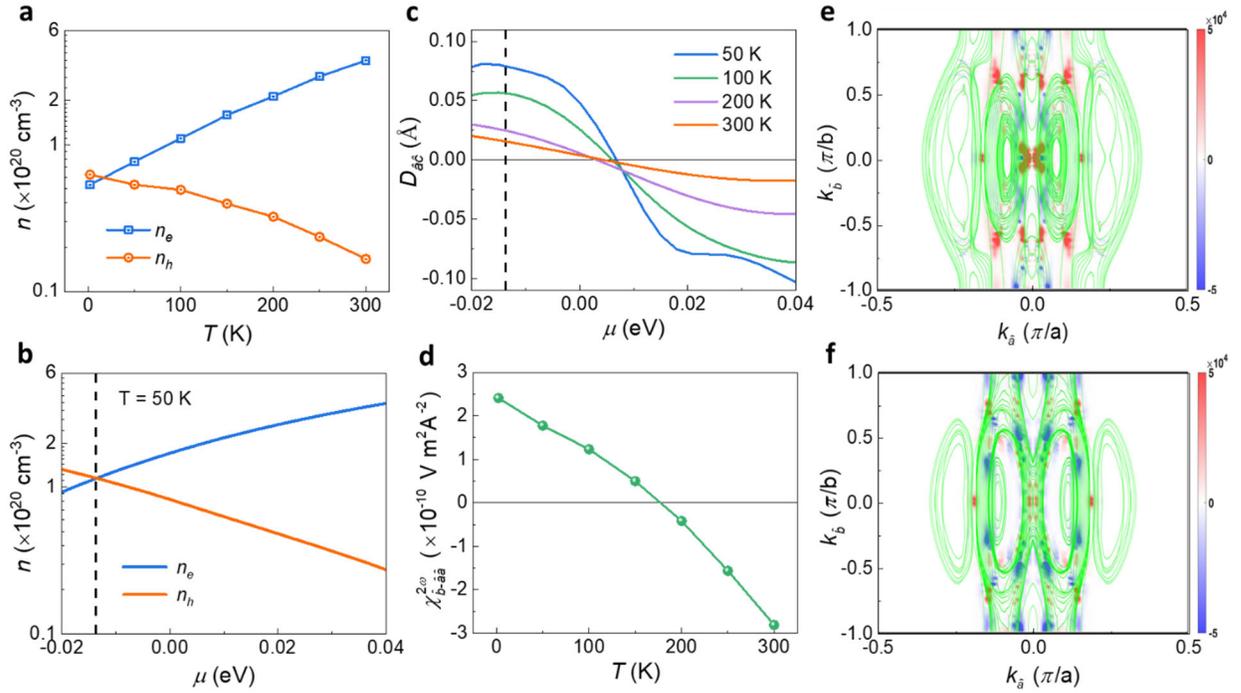

**Fig. 2 | Sign change of nonlinear Hall voltage due to the shift of chemical potential.** Experimental (**a**) and theoretical (**b**) carrier density of electron $n_e$ and hole $n_h$ at different temperature T and chemical potential $\mu_E$, respectively. Extracting from the comparison between **a** and **b**, an effective shift of $\mu_E$ in DFT bands is approximated from −0.014 eV (indicated by a dashed line in **b**) to +0.04 eV for T from 2 to 300 K. **c,** Calculated BCD $D_{\hat{a}\hat{c}}$ at different $\mu_E$, where the sign inversion is reproduced within the window of $\mu_E$ shifting. Dashed line is at $\mu_E = -0.014$ eV. Theoretically calculated BCD response in **c** agrees well with the experimentally observed sign inversion of the surface current density normalized NLH-signal $\chi$ by changing temperature in **d**. Calculated momentum k-resolved distribution of $D_{\hat{a}\hat{c}}$ at $\mu_E = -0.014$ eV (**e**) and $\mu_E = +0.03$ eV (**f**), where the green lines are the constant energy contours at the corresponding $\mu_E$. The smearing temperatures used for the calculations shown in **e** and **f** is 100 K.

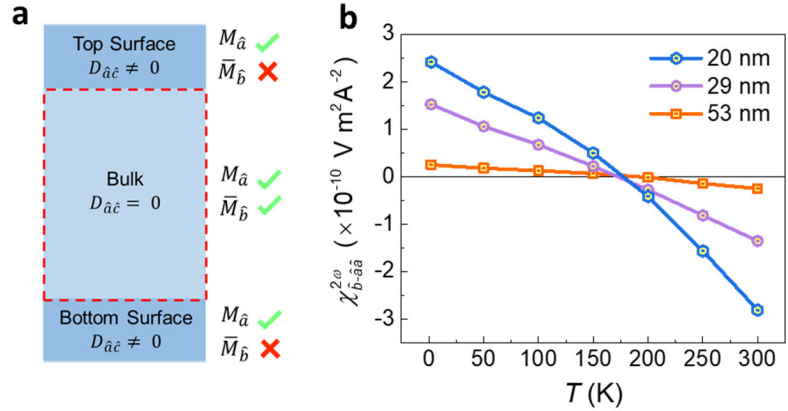

**Fig. 3 | Thickness dependent nonlinear Hall signal. a,** Schematic illustration for the contribution of BCD in a slab system with a large enough thickness. The slab can be decomposed into surface and bulk sectors. Red-cross-mark (green-right-mark) indicates the absence (presence) of mirror plane. Due to the presence of a glide mirror symmetry $\bar{M}_{\hat{b}}$, the bulk sector does not contribute to the NLH-signal and the suppression of $\chi$ is expected in thicker cases. Temperature evolution data of devices with different thickness is used to calculate the nonlinear Hall effect strength ($\chi$) plotted in **b**. All of the devices show a sign change of the nonlinear Hall signal, however, the values of $\chi$ is smaller for thicker devices.

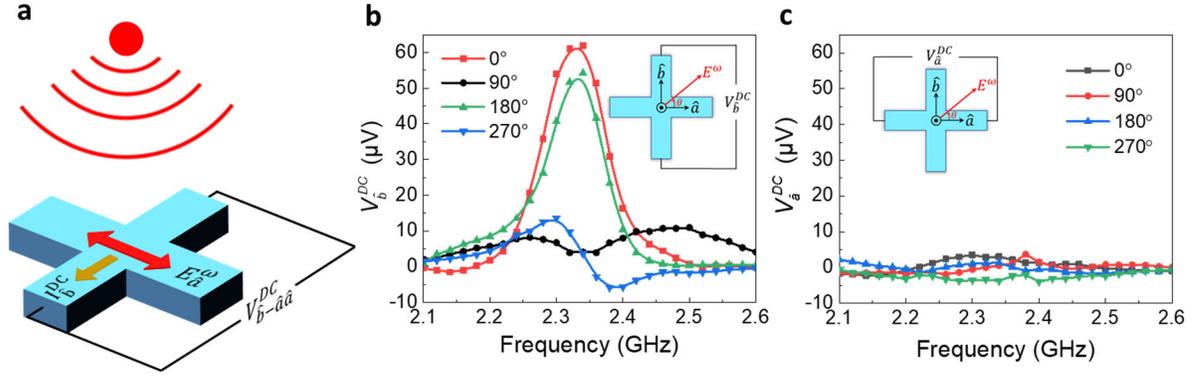

**Fig. 4 | Rectification demonstration based on NLHE in TaIrTe$_4$. a,** Schematic of the rectifier fabricated on 40 nm thick $T_d$-TaIrTe$_4$. Patch antenna with a high directivity and gain of 3 dBi at ~2.4 GHz is used for electromagnetic radiation. The Hall bar device, utilizing the room-temperature NLHE in TaIrTe$_4$, rectifies the incident electric field along $\hat{a}$ ($E_{\hat{a}}^{\omega}$) and generates a d.c. current along the transverse direction $\hat{b}$ ($I_{\hat{b}}^{DC}$). **b,c,** The rectified d.c. voltage measured along $\hat{b}$ ($V_{\hat{b}}^{DC}$) (**b**) and along $\hat{a}$ (**c**) as a function of frequency while aligning the incident electric field at four different angles (θ = 0°, 90°, 180°, and 270°) relative to $\hat{a}$ as shown in the inset. Maximum rectification occurs when the electric field is parallel to $\hat{a}$ (θ = 0°) and the voltage is measured along $\hat{b}$, in consistent with the symmetry constraints of NLHE.